\documentclass[10pt]{iopart}

\usepackage{braket}
\renewcommand{\vec}[1]{\mathbf{#1}}
\usepackage{graphicx}

\begin{document}
\bibliographystyle{iopart-num}
\title[\JPCM]{On the breakdown of the Born-Oppenheimer approximation in LiH and LiD}

\author{Ville J. H\"{a}rk\"{o}nen}

\address{Computational Physics Laboratory, Tampere University, P.O. Box 692, FI-33014 Tampere, Finland}
\ead{ville.j.harkonen@gmail.com}


\vspace{10pt}
\begin{indented}
\item[]January 2026
\end{indented}

\begin{abstract}
We compute the ab-initio electron density beyond the strict Born-Oppenheimer approximation in crystalline LiH and LiD with density functional methods. By taking into account the quantum mechanical nature of the nuclei, an aspect absent in the strict Born-Oppenheimer approximation, we find significant corrections to electron density in the vicinity of nuclei equilibrium positions. We compare our results with earlier experimental findings that have suggested a breakdown of the Born-Oppenheimer approximation in these solids and obtain improved agreement between experiment and theory when quantum nuclear effects are included. A notable temperature dependence of electron density is found. The results indicate the existence of beyond strict Born-Oppenheimer effects in solids at normal pressures and suggest that such effects can be significant also in materials containing light elements other than hydrogen.
\end{abstract}

%
\vspace{2pc}
\submitto{\JPCM}
%
%
\ioptwocol

\section{Introduction}
\label{Introduction}

The cornerstone of our current understanding of molecules and solids is the Born-Oppenheimer (BO) approximation \cite{Born-Oppenheimer-Adiabatic-Approx.1927,Born-Huang-DynamicalTheoryOfCrystalLattices-1954}, which makes the full quantum mechanical electron-nuclear Coulomb many-body problem computationally more feasible. It relies on the mass difference between nuclei and electrons and has proven to hold for a great majority of molecules and solids. The validity of the methods like the conventional density functional theory (DFT) \cite{Hohenberg-DFT-PhysRev.136.B864-1964,KohnSham-DFT-PhysRev.140.A1133-1965,DreizlerGross-DFTbook-1990} and the electronic many-body Green's function theory \cite{Gross-ManyParticleTheory-1991} is even more restrictive. Namely, these approaches assume the so-called strict BO approximation \cite{Izmaylov-EntanglementInTheBornOppenheimerApproximation-2017} (sometimes also called the crude BO approximation), the essence of which is that all the electronic quantities are strictly obtained from the electronic BO equation alone. In the strict BO approximation, the nuclear variables are parameters and the nuclear masses do not have any effect on the electronic structure. A well documented example of the BO breakdown is LiH molecule, which has been studied by experimental \cite{Wharton-PreliminaryValuesOfSomeMolecularConstantsOfLithiumHydride-1962} and computational \cite{Cafiero-NonBornOppenheimerCalculationsOfAtomsAndMolecules-2003,Bubin-AnAccurateNonBornOppenheimerCalculationOfTheFirstPurelyVibrationalTransitionInLiHmolecule-2005,Bubin-NonBornOppenheimerVariationalCalculationsOfAtomsAndMoleculesWithExplicitlyCorrelatedGaussianBasisFunctions-2005} methods.

The X-ray diffraction experiments conducted around 30 years ago, measuring electron density, suggest the breakdown of the BO approximation also in crystalline LiH \cite{Vidal-EvidenceOnTheBreakdownOfTheBornOppenheimerApproximationInTheChargeDensityOfCrystalline7LiHD-1992}. There have also been subsequent theoretical, computational and experimental studies suggesting the breakdown of the BO approximation in various hydrides \cite{Gidopoulos-BreakdownOfTheBornOppenheimerDescriptionExplainsNeutronComptonScatteringAnomaly-PhysRevB.71.054106-2005,Krzystyniak-AbInitioNuclearMomentumDistributionsInLithiumHydrideAssessingNonadiabaticEffects-PhysRevB.83.134305-2011,Krzystyniak-MassSelectiveNeutronSpectroscopyOfLithiumHydrideAndDeuterideExperimentalAssessmentOfTheHarmonicAndImpulseApproximations-PhysRevB.88.184304-2013}. Such a breakdown in hydrides would be relevant due to the high-temperature superconductivity recently discovered in these systems \cite{Xie-SuperconductivityOfLithiumDopedHydrogenUnderHighPressure-2014,Drozdov-ConventionalSuperconductivityAt203KelvinAtHighPressuresInTheSulfurHydrideSystem-2015,Drozdov-SuperconductivityAt250KInLanthanumHydrideUnderHighPressures-2019,Somayazulu-EvidenceForSuperconductivityAbove260KInLanthanumSuperhydrideAtMegabarPressures-PhysRevLett.122.027001-2019}. Moreover, hydrides have potential in hydrogen storage \cite{Ichikawa-MechanismOfNovelReactionFromLiNH2AndLiHtoLi2NHandH2asAPromisingHydrogenStorageSystem-2004,Klopvcivc-AreviewOnMetalHydrideMaterialsForHydrogenStorage-2023}. Ab-initio computations of electronic \cite{Baroni-QuasiparticleBandStructureOfLithiumHydride-PhysRevB.32.4077-1985,vanSetten-ElectronicStructureAndOpticalPropertiesOfLightweightMetalHydrides-PhysRevB.75.035204-2007,Reshak-MgH2andLiHmetalHydridesCrystalsAsNovelHydrogenStorageMaterialElectronicStructureAndOpticalProperties-2013} and lattice dynamical properties \cite{Biswas-AbInitioStudyOfTheLiHphaseDiagramAtExtremePressuresAndTemperatures-PhysRevB.99.024108-2019} of LiH have been established. While the effect of quantum mechanical nuclei has been studied on some properties of hydrogen rich systems \cite{Scivetti-EffectOfQuantizationOfVibrationsOnTheStructuralPropertiesOfCrystals-PhysRevB.78.224108-2008,Monacelli-QuantumPhaseDiagramOfHighPressureHydrogen-2023,vanDeBund-CompetitionBetweenSuperconductivityAndMolecularizationInTheQuantumNuclearBehaviorOfLanthanumHydride-PhysRevB.108.184102-2024}, there are no earlier computational studies which would have taken the quantum nuclear effects into account in the study of electron density in solids. In particular, there has been no computational studies trying to explain the earlier experimental results suggested breakdown of the BO approximation in crystalline LiH and LiD \cite{Vidal-EvidenceOnTheBreakdownOfTheBornOppenheimerApproximationInTheChargeDensityOfCrystalline7LiHD-1992}. Here we establish the first ab-initio computation of electron density beyond the strict BO approximation in crystalline LiH and LiD with density functional methods and compare our result to the earlier experimental findings \cite{Vidal-EvidenceOnTheBreakdownOfTheBornOppenheimerApproximationInTheChargeDensityOfCrystalline7LiHD-1992}. We develop two different approaches to compute the elecrtron density and thus X-ray structure factors with quantum mechanical nuclei. 

This work is organized as follows. We summarize the theoretical background in Sec. \ref{Theory} and develop two different approaches to compute electron densities with quantum mechanical nuclei in Secs. \ref{HarmonicApproximationToDensity} and \ref{GaussianApproachToDensity}. We give the calculational details in Sec. \ref{CalculationalDetails}. The computed electron densities are represented in Sec. \ref{ElectronDensity} and we make a comparison with the experimental results in Secs. \ref{ComparisonWithExperimentalRadialDensities} and \ref{ComparisonWithExperimentalStructureFactors}.

\section{Theory}
\label{Theory}

\subsection{Background}
\label{Background}

The starting point of the approach used here is the many-body Coulomb problem of electrons and nuclei described by the Coulomb Hamiltonian $H$. The solution of the Schr\"{o}dinger equation $H \Psi\left(\vec{r},\vec{R}\right) = E \Psi\left(\vec{r},\vec{R}\right)$ is not computationally feasible, in general. In the seminal work by Born and Oppenheimer \cite{Born-Oppenheimer-Adiabatic-Approx.1927}, it was shown that the exact problem can be separated into two parts: one for the electrons and one for the nuclei \cite{Born-Huang-DynamicalTheoryOfCrystalLattices-1954}
\begin{eqnarray} 
H_{n} \chi\left(\vec{R}\right) &= E \chi\left(\vec{R}\right), \label{eq:TheoryEq_0} \\
H_{BO} \Phi_{\vec{R}}\left(\vec{r}\right) &= \epsilon_{BO}\left(\vec{R}\right) \Phi_{\vec{R}}\left(\vec{r}\right), \label{eq:TheoryEq_1}
\end{eqnarray}
where the nuclear ($H_{n})$ and electronic BO Hamiltonians ($H_{BO}$) are of the form
\begin{equation} 
H_{BO} = H - T_{n}, \quad H_{n} = T_{n} + \epsilon_{BO}.
\label{eq:TheoryEq_2}
\end{equation}
Here, $T_{n}$ is the nuclear kinetic energy and $\epsilon_{BO}$ the BO energy from Eq. \ref{eq:TheoryEq_1}. Now, many approaches, including DFT \cite{Hohenberg-DFT-PhysRev.136.B864-1964,KohnSham-DFT-PhysRev.140.A1133-1965} as implemented in program packages like Quantum Espresso (QE) \cite{Giannozzi-QuantumEspresso-2009} and in Green's function theory \cite{Gross-ManyParticleTheory-1991}, one actually considers the conditional BO electron density obtainable from Eq. \ref{eq:TheoryEq_1}, only. In some systems this strict BO approximation: Eqs. \ref{eq:TheoryEq_0} and \ref{eq:TheoryEq_1} as separate entitiesd do not necessarily give accurate results.

Over the years several alternative beyond-BO approaches have been developed \cite{Kreibich-MulticompDFTForElectronsAndNuclei-PhysRevLett.86.2984-2001,Gidopoulos-ElectronicNonAdiabaticStates-2005,Gidopoulos-Gross-ElectronicNonAdiabaticStates-2014,Abedi-ExactFactorization-PhysRevLett.105.123002-2010,Villaseco-ExactFactorizationAdventuresAPromisingApproachForNonBoundStates-2022,Muolo-NuclearElectronicAllParticleDensityMatrixRenormalizationGroup-2020,Feldmann-QuantumProtonEffectsFromDensityMatrixRenormalizationGroupCalculations-2022} since the beyond-BO wave function approach is not computationally feasible in numerous cases, including solids. One of these alternative approaches is the beyond-BO many-body Green's function theory \cite{Baym-field-1961,Harkonen-ManyBodyGreensFunctionTheoryOfElectronsAndNucleiBeyondTheBornOppenheimerApproximation-PhysRevB.101.235153-2020,Harkonen-ExactFactorizationOfTheManyBodyGreensFunctionTheoryOfElectronsAndNuclei-PhysRevB.106.205137-2022,Stefanucci-InAndOutOfEquilibriumAbInitioTheoryOfElectronsAndPhonons-PhysRevX.13.031026-2023,Harkonen-QuantumFieldTheoryOfElectronsAndNuclei-2024,Harkonen-CorrigendumQuantumFieldTheoryOfElectronsAndNuclei-2024J.Phys.A.Math.Theor.57.465402-2025}, which allows the computation of exact observables like electron density. As with the wave function approach, the general form of the Green's function theory is not computationally feasible and to overcome this limitation we combined the exact factorization  \cite{Hunter-ConditionalProbInWaveMech-1975,Gidopoulos-ElectronicNonAdiabaticStates-2005,Gidopoulos-Gross-ElectronicNonAdiabaticStates-2014,Abedi-ExactFactorization-PhysRevLett.105.123002-2010,Abedi-CorrelatedElectronNuclearDynamicsEF-2012,Schild-ElectronicFluxDensityBeyondTheBornOppenheimerApproximation-2016,Agostini-ExactFactorizationOfTheElectronNuclearWaveFunctionTheoryAndApplications-2020} and Green's function approaches \cite{Harkonen-ExactFactorizationOfTheManyBodyGreensFunctionTheoryOfElectronsAndNuclei-PhysRevB.106.205137-2022}. The exact factorization of the Green's function theory \cite{Harkonen-ExactFactorizationOfTheManyBodyGreensFunctionTheoryOfElectronsAndNuclei-PhysRevB.106.205137-2022} provides a systematic way for deriving approximations beyond-BO. In particular, it can be shown by using the exact factorization of the Green's functions that \cite{Harkonen-ExactFactorizationOfTheManyBodyGreensFunctionTheoryOfElectronsAndNuclei-PhysRevB.106.205137-2022} $G\left(\vec{y}t,\vec{y}'t'\right) \approx \int d\vec{R} \rho\left(\vec{R}\right) G_{\vec{R}}\left(\vec{y}t,\vec{y}'t'\right)$ and thus an approximate beyond-BO approximation to electron density can be written as
\begin{equation} 
\left\langle n\left(\vec{y}\right)\right\rangle = \int d\vec{R} \rho\left(\vec{R}\right) n_{\vec{R}}\left(\vec{y}\right),
\label{eq:TheoryEq_3}
\end{equation}
where $\rho\left(\vec{R}\right)$ is the many-body nuclear density, including the temperature dependent case (canonical ensemble for the nuclei). A special case of this is the zero temperature form $\rho\left(\vec{R}\right) = \left|\chi\left(\vec{R}\right)\right|^{2}$ that can be obtained directly from the exact factorization approach \cite{Gidopoulos-ElectronicNonAdiabaticStates-2005,Gidopoulos-Gross-ElectronicNonAdiabaticStates-2014}. Notably, a further special case is when $n_{\vec{R}}\left(\vec{y}\right)$ is the BO electron density extractable from Eq. \ref{eq:TheoryEq_1} and $\chi\left(\vec{R}\right)$ the nuclear wave function satisfying Eq. \ref{eq:TheoryEq_0}. In this particular case, referring to the work of Huang and Born \cite{Born-Huang-DynamicalTheoryOfCrystalLattices-1954}, we are actually computing electron density within the BO approximation. However, this is not the same density as computed in DFT and Green's function theories within the BO approximation. Instead, in these approaches we are computing the conditional BO electron density $n_{\vec{R}}\left(\vec{y}\right)$. To distinct the differences in these electron densities we name them as follows. The density from Eq. \ref{eq:TheoryEq_3}, $n\left(\vec{y}\right)$, when the BO approximation is assumed in the computation of $\chi\left(\vec{R}\right)$ and $n_{\vec{R}}\left(\vec{y}\right)$, is called the electron density in the full BO approximation. In this situation, we further call $n_{\vec{R}}\left(\vec{y}\right)$ the conditional electron density in the strict BO approximation. The integral in Eq. \ref{eq:TheoryEq_3} is demanding to compute since we have to solve $n_{\vec{R}}\left(\vec{y}\right)$ for all nuclear configurations that the integral goes through. In the following, we introduce approximate approaches to compute the electron density $n\left(\vec{y}\right)$ from Eq. \ref{eq:TheoryEq_3}.

\subsection{Harmonic Approximation to Density}
\label{HarmonicApproximationToDensity}

In the following we derive an approximate way of solving Eq. \ref{eq:TheoryEq_3} and see how the temperature dependence of Eq. \ref{eq:TheoryEq_3} originates by first considering the system in a nuclear eigenstate within the harmonic approximation, $\chi_{m}\left(\vec{R}\right)$. We can solve the BO nuclear equation in the harmonic approximation, which holds for materials with weak anharmonicity, to solve an exact nuclear wave function $\chi_{m}\left(\vec{R}\right)$. The procedure is as follows. We carry out the conventional coordinate transformation of lattice dynamics \cite{Born-Huang-DynamicalTheoryOfCrystalLattices-1954} $\vec{R} = \vec{x} + \vec{u}$, where $\vec{x}$ are parameters called the reference positions and $\vec{u}$ are the displacements (quantum mechanical variables) from the reference positions. We expand $n_{\vec{R}}\left(\vec{y}\right) = n_{\vec{x} + \vec{u}}\left(\vec{y}\right)$ up to third order in $\vec{u}$ about $\vec{x}$ and a special case of Eq. \ref{eq:TheoryEq_3}, $n_{m}\left(\vec{y}\right) = \int d\vec{R} \left|\chi_{m}\left(\vec{R}\right)\right|^{2} n_{\vec{R}}\left(\vec{y}\right)$, can be approximated as
\begin{equation} 
n_{m}\left(\vec{y}\right) \approx n_{\vec{x}}\left(\vec{y}\right) + \frac{1}{2} \sum_{s, s'} \frac{\partial^{2}{n_{\vec{x}}\left(\vec{y}\right)}}{\partial{x_{s}} \partial{x_{s'}}} \braket{m|\hat{u}_{s} \hat{u}_{s'}|m},
\label{eq:TheoryEq_4}
\end{equation}
where $\ket{m}$ is the $m$th eigenstate of the harmonic nuclear Hamiltonian. Here we used the fact that the first and third order terms vanish since we assumed the harmonic eigenstates. In Eq. \ref{eq:TheoryEq_4}, $s = \alpha k$ is the combined label for Cartesian index $\alpha$ and the label for the $k$th nucleus. We can incorporate temperature dependence by taking harmonic canonical thermal average of Eq. \ref{eq:TheoryEq_4} with respect to the nuclear states, namely
\begin{equation}
\left\langle n\left(\vec{y}\right)\right\rangle = n_{\vec{x}}\left(\vec{y}\right) + \frac{1}{2} \sum_{s, s'} \frac{\partial^{2}{n_{\vec{x}}\left(\vec{y}\right)}}{\partial{x_{s}} \partial{x_{s'}}} \left\langle \hat{u}_{s} \hat{u}_{s'} \right\rangle.
\label{eq:TheoryEq_5}
\end{equation}
We denote the last term of Eq. \ref{eq:TheoryEq_5} as $\left\langle n'\left(\vec{y}\right)\right\rangle \equiv \frac{1}{2} \sum_{s, s'} \frac{\partial^{2}{n_{\vec{x}}\left(\vec{y}\right)}}{\partial{x_{s}} \partial{x_{s'}}} \left\langle \hat{u}_{s} \hat{u}_{s'} \right\rangle$ and use the following notation for the ensemble average
\begin{equation} 
\left\langle \hat{o} \right\rangle = \sum_{m} p_{m} \braket{m|\hat{o}|m}, \quad p_{m} = \frac{e^{-\beta E_{m}}}{\sum_{m'} e^{-\beta E_{m'}}},
\label{eq:TheoryEq_6}
\end{equation}
where $E_{m}$ is the eigenvalue of the $m$th eigenstate of the harmonic nuclear Hamiltonian. The quantities needed to compute $\left\langle n\left(\vec{y}\right)\right\rangle$ are therefore the equilibrium BO electron density $n_{\vec{x}}\left(\vec{y}\right)$, its second order mixed partial derivatives and the covariance
\begin{equation} 
\left\langle \hat{u}_{s} \hat{u}_{s'}\right\rangle = \frac{ \hbar }{ \sqrt{ M_{s} M_{s'} } } \sum_{j} \frac{e\left(s|j\right) e\left(s'|j\right)}{\omega_{j}} \left(\bar{n}_{j} + \frac{1}{2}\right),
\label{eq:TheoryEq_7}
\end{equation}
where $e\left(s|j\right)$ is the eigenvector of the normal modes, $\omega_{j}$ the normal mode frequency and $\bar{n}_{j} = {1}/{(e^{\beta \hbar \omega_{j}} - 1)}$ the Bose-Einstein distribution function. The relation of Eq. \ref{eq:TheoryEq_7} takes a similar form when the phonon coordinates are used \cite{maradudin-harm-appr-1971,Harkonen-NTE-2014} and both ways will give the same values of $\left\langle \hat{u}_{s} \hat{u}_{s'}\right\rangle$. We compute these quantities by using open source ab-initio computational package QE \cite{Giannozzi-QuantumEspresso-2009}, which is based on DFT \cite{Hohenberg-DFT-PhysRev.136.B864-1964,KohnSham-DFT-PhysRev.140.A1133-1965}. The QE program package has been successfully used to predict various experimentally relevant quantities of interest \cite{Giannozzi-AdvancedCapabilitiesForMaterialsModellingWithQuantumESPRESSO-2017} within BO approximation, including phonon related properties \cite{Harkonen-Tcond-II-VIII-PhysRevB.93.024307-2016,Harkonen-AbInitioComputationalStudyOnTheLatticeThermalConductivityOfZintlClathrates-PhysRevB.94.054310-2016,Tadano-FirstPrinciplesPhononQuasiparticleTheoryAppliedToAStronglyAnharmonicHalidePerovskite-PhysRevLett.129.185901-2022}.

\subsection{Gaussian Approach to Density}
\label{GaussianApproachToDensity}

We note that results similar to the ensemble-averaged density in Eq. \ref{eq:TheoryEq_5} have been considered previously for molecules, in a approach known as thermal or vibrational smearing of the electron density \cite{Coulson-TheEffectOfMolecularVibrationsOnApparentBondLengths-1971,Hirshfeld-XIIIChargeDeformationAndVibrationalSmearing-1977,Michael-ValidationOfConvolutionApproximationToTheThermalAverageElectronDensity-2015}. In these approaches, no Taylor expansion in the nuclear variables is established, but instead the so-called convolution approximation \cite{Coulson-TheEffectOfMolecularVibrationsOnApparentBondLengths-1971,Hirshfeld-XIIIChargeDeformationAndVibrationalSmearing-1977,Michael-ValidationOfConvolutionApproximationToTheThermalAverageElectronDensity-2015,Hartikainen-TheEffectOfTheQuantumNatureOfNucleiOnTheElectronDensityOfCrystallineSolids-2025} is imposed simplifying the calculation of the thermally averaged version of Eq. \ref{eq:TheoryEq_3}. Here we derive an alternative approach to compute the density from Eq. \ref{eq:TheoryEq_3} based on Gaussian function fitted strict BO electron density and find it useful in particular in the calculation of all electron densities.

In this approach, we write the strict BO density in terms of Gaussian basis functions as
\begin{equation} 
n_{\vec{R}}\left(\vec{y}\right) = \sum_{k,l,b} c_{klb}\left(\vec{R}\right) \chi_{klb}\left(\vec{y}\right),
\label{eq:GaussianApproachToDensityEq_1}
\end{equation}
where $c_{kls}\left(\vec{R}\right)$ are coefficients to be found, $l$ label the angular momenta state, $b$ different parameters in the expansion and \cite{Helgaker-MolecularElectronicStructureTheory-2000}
\begin{equation} 
\chi_{klb}\left(\vec{y}\right) = \prod^{3}_{\alpha = 1} \left[y_{\alpha} - R_{\alpha}\left(k\right)\right]^{l_{\alpha}} e^{- \xi_{klb} \left[y_{\alpha} - R_{\alpha}\left(k\right)\right]^{2} }.
\label{eq:GaussianApproachToDensityEq_2}
\end{equation}
There are few benefits of doing such a fit and such a decomposition is essentially always assumed when measured structure factors are used to refinement procedure of the experimental electron density \cite{Vidal-EvidenceOnTheBreakdownOfTheBornOppenheimerApproximationInTheChargeDensityOfCrystalline7LiHD-1992}. Given there is a relatively weak dependence of $c_{klb}\left(\vec{R}\right)$ on $\vec{R}$, the integral of Eq. \ref{eq:TheoryEq_3} becomes much easier to calculate as only one nuclear coordinate appears in the integral for each nuclei $k$. Also, there are well known results for Gaussian integrals allowing us to derive a closed form expressions for the remaining integrals. In this case we denote $\left\langle n\left(\vec{y}\right)\right\rangle \approx \left\langle n_{G}\left(\vec{y}\right)\right\rangle$ and an approximate for of Eq. \ref{eq:TheoryEq_3} can be written as
\begin{eqnarray} 
\left\langle n_{G}\left(\vec{y}\right)\right\rangle =& \sum_{k,l,b} \prod^{3}_{\alpha = 1} c_{klb} \int d\vec{R}_{k} \rho_{k}\left(\vec{R}_{k}\right) \nonumber \\
&\times \left[y_{\alpha} - R_{\alpha}\left(k\right)\right]^{l_{\alpha}} e^{- \xi_{klb} \left[y_{\alpha} - R_{\alpha}\left(k\right)\right]^{2} }.
\label{eq:GaussianApproachToDensityEq_3}
\end{eqnarray}
In the harmonic approximation
\begin{equation} 
\rho_{k}\left[R_{\alpha}\left(k\right)\right] = \left(\frac{1}{2 \pi \sigma^{2}_{\alpha k}}\right)^{1/2} e^{- \frac{\left[R_{\alpha}\left(k\right) - x_{\alpha}\left(k\right)\right]^{2}}{2 \sigma^{2}_{\alpha k}} },
\label{eq:GaussianApproachToDensityEq_4}
\end{equation}
where $\sigma^{2}_{\alpha k} = \langle \hat{u}^{2}_{\alpha}\left(k\right) \rangle$ and Eq. \ref{eq:GaussianApproachToDensityEq_3} can be written as
\begin{eqnarray} 
\left\langle n_{G}\right\rangle =& \sum_{k,l,b} c_{klb} \prod^{3}_{\alpha = 1} \left(\frac{1}{2 \pi \sigma^{2}_{\alpha k}}\right)^{1/2} \sqrt{\frac{\pi}{A_{\alpha klb}}} \frac{i^{-l_{\alpha}}}{2^{l_{\alpha}} A^{l_{\alpha}/2}_{\alpha klb}}  \nonumber \\
&\times H_{l_{\alpha}}\left[i(y_{\alpha}-\mu_{\alpha klb})\sqrt{A_{\alpha klb}}\right] \exp\left[ D_{\alpha klb} \right].
\label{eq:GaussianApproachToDensityEq_5}
\end{eqnarray}
Here $H_{l_{\alpha}}\left[z\right]$ are physicist Hermite polynomials and
\begin{eqnarray}
A_{\alpha klb} =& \xi_{klb} + \frac{1}{2\sigma_{\alpha k}^{2}}, \nonumber \\
\mu_{\alpha klb} =& \frac{2\xi_{klb} y_{\alpha} + x_{\alpha}(k)/\sigma_{\alpha k}^{2}}{2\xi_{klb} + 1/\sigma_{\alpha k}^{2}}, \nonumber \\
D_{\alpha klb} =& -\xi_{klb} y_{\alpha}^{2} - \frac{x^{2}_{\alpha}(k)}{2\sigma_{\alpha k}^{2}} \nonumber \\
&+ \frac{\left[2\xi_{klb} y_{\alpha} + x_{\alpha}(k)/\sigma_{\alpha k}^{2}\right]^{2}}{4 \left[\xi_{klb}+1/(2\sigma_{\alpha k}^{2})\right]}.
\label{eq:GaussianApproachToDensityEq_6}
\end{eqnarray}
Indeed, it can be shown by using the polynomial approach of Sec. \ref{HarmonicApproximationToDensity} that the main assumption of the approach derived here (electron density as a sum of single atom contributions) likely holds to a rather high degree of accuracy in the studied system. This is also in line with the earlier works on molecules using the convolution approximation \cite{Coulson-TheEffectOfMolecularVibrationsOnApparentBondLengths-1971,Hirshfeld-XIIIChargeDeformationAndVibrationalSmearing-1977,Michael-ValidationOfConvolutionApproximationToTheThermalAverageElectronDensity-2015} a particular case of which the presented Gaussian approach also is.

\section{Calculational details}
\label{CalculationalDetails}

Calculations were performed using Quantum Espresso (version 7.3.1), employing the projector augmented-wave (PAW) method \cite{Blochl-ProjectorAugmentedWaveMethod-PhysRevB.50.17953-1994} with both the PBE \cite{Perdew-GeneralizedGradientApproximationMadeSimple-PhysRevLett.77.3865-1996} and LDA \cite{KohnSham-DFT-PhysRev.140.A1133-1965,Perdew-SelfInteractionCorrectionToDensityFunctionalApproximationsForManyElectronSystems-PhysRevB.23.5048-1981} exchange–correlation functionals, and PAW datasets from the PSLibrary (PBE calculations) \cite{dalCorso-PseudopotentialsPeriodicTableFromHtoPu-2014} and JTH library (v2.0, LDA calculations) \cite{Jollet-GenerationOfProjectorAugmentedWaveAtomicDataA71ElementValidatedTableInTheXMLFormat-2014}. The harmonic phonon and normal mode frequencies needed in Eq. \ref{eq:TheoryEq_7} are computed by using the density functional perturbation theory \cite{Gonze-DynMatBornEffChargesDielectricPermiTensorsAndIFCsFromDFPT-PhysRevB.55.10355-1997} as implemented in QE \cite{Baroni-PhononsAndRelatedCrystalPropFromDFTPT-RevModPhys.73.515-2001}. As we consider an ionic crystal, the long-range force constants are included and the non-analytic part of the dynamical matrix is used at $\vec{q} = 0$. The plane wave kinetic energy and charge density cut-off values used were 100 Ry and 1200 Ry, respectively. The electronic structure was computed with $16 \times 16 \times 16$ $\vec{k}$ point grids. We constructed $2 \times 2 \times 2$ supercells in order to compute the electron density derivatives of Eq. \ref{eq:TheoryEq_5} and the necessary related lattice dynamical properties. The phonon dispersion relations and the covariances used with the Gaussian approach were computed with $8 \times 8 \times 8$ $\vec{q}$ point grid. The derivatives were computed as finite central differences with 0.5\% displacements from the nuclear equilibrium positions. To compute the electron density corrections from Eq. \ref{eq:TheoryEq_5} we computed the lattice dynamical properties with the $\vec{q}$ point meshes matching the supercell dimensions. The structure parameters for LiH ($Fm\bar{3}m$), which define the structure, are given in Table 1 of \cite{Vidal-EvidenceOnTheBreakdownOfTheBornOppenheimerApproximationInTheChargeDensityOfCrystalline7LiHD-1992}. The LiH structure parameters used are the following: lattice parameter $a = 7.674$~au (bohr); fractional coordinates of the in-equivalent atoms Li$ = \left(0.000, 0.000, 0.000\right)$, H$ = \left(0.500, 0.000, 0.000\right)$. We first established the structure relaxation of the structure with the given parameters after which the lattice dynamical properties were computed. After relaxation, we found lattice parameters $a = 7.564$~au for PBE and $a = 7.400$~au for LDA computations. The equilibrium structure of LiD is identical to LiH in the strict BO approximation. All the calculations are established at 0 kbar pressure.

\section{Results}
\label{Results}

\subsection{Electron density}
\label{ElectronDensity}

\begin{figure*}
\centering
\includegraphics[width=1.0\textwidth]{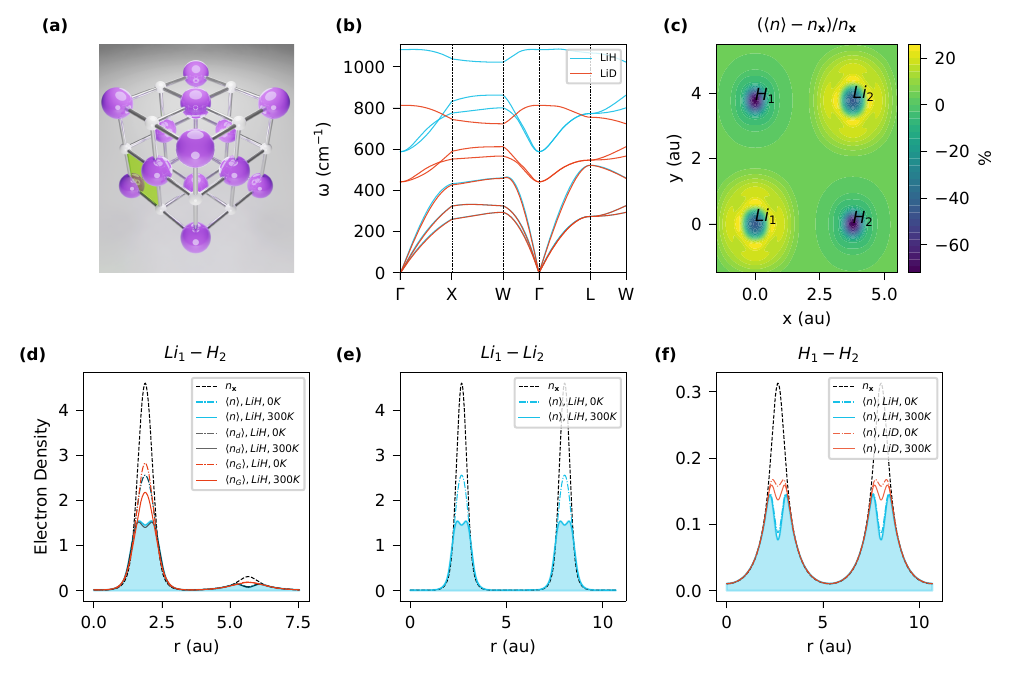}
\caption{Phonon dispersions for LiH and LiD and pseudo electron densities in 100-plane of the conventional unit cell (110-plane of the primitive cell) and in selected lines along the 100-plane. The electron densities are normalized to the number of electrons per unit cell. (\textbf{a}) The conventional unit cell with the 100-plane indicated with a green plane in the lower left corner (\textbf{b}) phonon dispersions for LiH and LiD (\textbf{c}) the difference of beyond strict BO and strict BO densities relative to $n_{\vec{x}}\left(\vec{y}\right)$ in percentage at 0 K, (\textbf{d}) electron densities along the line between lithium and hydrogen nuclei (pointed out in \textbf{c}), (\textbf{e}) between lithium nuclei and (\textbf{f}) between hydrogen nuclei. In (\textbf{d}), $\left\langle n_{d}\right\rangle$ denotes the diagonal contribution to $\left\langle n\right\rangle$ discussed in the text. We have left out the lines for LiD in (\textbf{d}) and (\textbf{e}) since in the vicinity of lithium nuclei the results are essentially identical.}
\label{fig1}
\end{figure*}
The conventional unit cell of the LiH crystal structure is given in Fig. \ref{fig1}(\textbf{a}) and the phonon dispersions for LiH and LiD in Fig. \ref{fig1}(\textbf{b}). The phonon dispersion closely resembles to that obtained in earlier studies \cite{Zhang-PhononAndElasticInstabilitiesInRocksaltAlkaliHydridesUnderPressureFirstPrinciplesStudy-PhysRevB.75.104115-2007,Biswas-AbInitioStudyOfTheLiHphaseDiagramAtExtremePressuresAndTemperatures-PhysRevB.99.024108-2019}. The acoustic modes of LiH and LiD are rather close to identical implying that these modes almost completely consist of vibrations of Li atoms. As expected, the optical modes in LiD are scaled down relative to the corresponding modes of LiH due to the higher mass of deuterium.

We next consider the results obtained with the polynomial approach. The relative change $\left[\left\langle n\left(\vec{y}\right)\right\rangle - n_{\vec{x}}\left(\vec{y}\right)\right] / n_{\vec{x}}\left(\vec{y}\right) = \left\langle n'\left(\vec{y}\right)\right\rangle / n_{\vec{x}}\left(\vec{y}\right)$ in pseudo electron density in the 100-plane of the conventional cell [see Fig. \ref{fig1}(\textbf{a})] is depicted in Fig. \ref{fig1}(\textbf{c}). Moreover, pseudo electron densities along different lines of the 100-plane are given by Figs. \ref{fig1}(\textbf{d})-\ref{fig1}(\textbf{f}). We see a significant breakdown of the strict BO approximation. The largest relative change is around -76\%/-82\% (0 K/300 K) at the hydrogen nuclear equilibrium positions and around -47\%/-75\% (0 K/300 K) at the lithium equilibrium positions. The largest positive relative change is in the surrounding volumes of the Li nuclear positions. The relative change is positive in the large volumes between the nuclei. In the case of LiD, the largest relative change is around -53\%/-62\% (0 K/300 K) at the deuterium nuclear equilibrium positions, the changes around Li nuclei being essentially identical to that in LiH. The breakdown is more significant at higher temperatures, especially in the vicinity of the heavier nuclei, lithium and deuterium. We see drastic deviations at 300 K as the strict BO values are about five times higher than the full BO values at the equilibrium positions of hydrogen and lithium. The functional form of the full BO and strict BO densities at 0 K is the same around the Li nuclei, but around the hydrogen(deuterium) nuclei, the change from unimodal to bimodal functional shape occurs. At 300 K the functional forms of the densities near all nuclei, including lithium, change from unimodal to bimodal shape. In Fig. \ref{fig1}(\textbf{d}), the diagonal elements of Eq. \ref{eq:TheoryEq_5} are depicted and these terms mostly explain the full BO reduction of electron density with quantum mechanical nuclei. The effect of off-diagonal terms at the lithium nuclei equilibrium positions is less than 4\%. The change in electron density at a given point in space is thus mostly due to a local position uncertainty of the nucleus, which is the same mechanism that occurs in the YH$_{6}$ superconductor and Cs-IV phase of solid hydrogen \cite{Harkonen-BreakdownOfTheBornOppenheimerApproximationInSolidHydrogenAndHydrogenRichSolids-Arxiv-2023}. The diagonal form is $\left\langle n_{d}\left(\vec{y}\right)\right\rangle = n_{\vec{x}}\left(\vec{y}\right) + \frac{1}{2} \sum_{s} {\partial^{2}{n_{\vec{x}}\left(\vec{y}\right)}}/{\partial{x^{2}_{s}}} \left\langle \hat{u}^{2}_{s} \right\rangle$ and due to the fact that these terms mostly explain the smearing of electron density actually validates the convolution approximation \cite{Coulson-TheEffectOfMolecularVibrationsOnApparentBondLengths-1971,Hirshfeld-XIIIChargeDeformationAndVibrationalSmearing-1977,Michael-ValidationOfConvolutionApproximationToTheThermalAverageElectronDensity-2015} and thus also the approach developed in Sec. \ref{GaussianApproachToDensity}.
\begin{figure}
\centering
\includegraphics[]{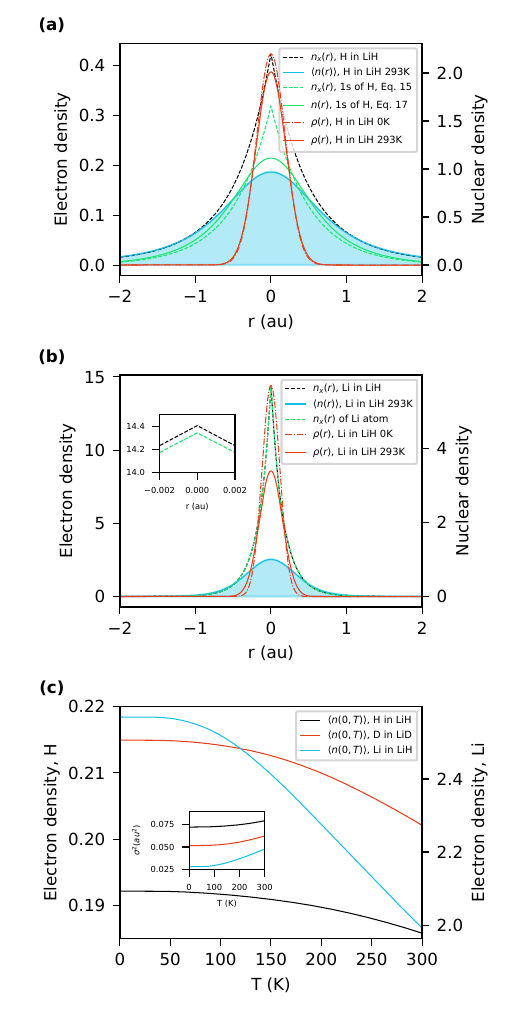}
\caption{Radial all electron densities by using Gaussian approach, nuclear densities, mean square displacements and electron density temperature dependence. (\textbf{a}) The nuclear one-body densities for hydrogen in LiH and the electron densities for hydrogen in LiH, the hydrogen atom, and the model of Eq. \ref{eq:ElectronDensity_Eq_3}. (\textbf{b}) The nuclear one-body densities for lithium in LiH and the electron densities for lithium in LiH and for the isolated lithium atom, solved numerically with the same DFT approach as in the crystalline case. For better visual, the inset shows the same curves near the nuclei equilibrium position. The axis for nuclear densities on the right. (\textbf{c}) The temperature dependence of electron densities at the nuclear equilibrium positions and mean square displacements in the inset, where the curve colors for elements matches those in the main plot.}
\label{fig2}
\end{figure}

The pseudo electron density computed by using Eq. \ref{eq:GaussianApproachToDensityEq_5} is depicted in Fig. \ref{fig1}(\textbf{d}). The results are similar to those obtained with the polynomial approach, but the electron density values near the nuclei equilibrium positions are slightly higher at both temperatures. Also the bimodal shape of the density is absent. The polynomial approximation is more rigorous than the Gaussian approach in the sense that the effect of all nuclei is included. The Gaussian approach is on the other hand more rigorous in the sense that no finite order expansion is assumed. Efforts in computing all electron density reveal a weakness of the polynomial approach. In short, we obtain large negative values of the full-BO densities with polynomial approach at the nuclei equilibrium positions, which cannot be solely originate from numerical errors and thus in the case of all electron densities non-physical results are obtained. These aspects may, at least partially, explain the bimodal shapes visible in the pseudo electron density and originate from the cusps in the strict-BO all electron densities at the nuclei equilibrium positions.

We discuss these issues of the polynomial approach, in the case of all electron density, through the following example. First, the form of the electron density in displacement of any nuclei remains approximately invariant. This sets a constraint on a functional form and we illustrate this by considering the electron density of the lowest energy zero angular momentum state of hydrogen, the wave function can be written as
\begin{equation} 
n_{R}\left(r\right) = \frac{1}{\pi a^{3}_{0}} \exp\left[-2\frac{\left|r - R\right|}{a_{0}}\right],
\label{eq:ElectronDensity_Eq_1}
\end{equation}
and a plot of this function at $R = x = 0$ is depicted in Fig. \ref{fig2}(\textbf{a}). Consider now a simple model of Eq. \ref{eq:TheoryEq_3}
\begin{equation} 
n\left(r\right) \equiv \int^{\infty}_{-\infty} dR \left|\chi\left(R\right)\right|^{2} n_{R}\left(r\right),
\label{eq:ElectronDensity_Eq_2}
\end{equation}
where $\left|\chi\left(R\right)\right|^{2} \equiv {1}/{\sqrt{2 \pi \sigma}} \exp\left[- \frac{1}{2} {\left(R - x\right)^2}/{\sigma^2}\right]$ as in our general harmonic case. The integral of Eq. \ref{eq:ElectronDensity_Eq_2} computes to
\begin{eqnarray} 
n\left(r\right) =& \frac{e^{2 \sigma^{2}/a^{2}_{0}}}{\pi a^{3}_{0}} \left[ e^{-2 \left(r - x\right)/a_{0}} \Phi\left(\frac{r - x}{\sigma} - \frac{2 \sigma}{a_{0}}\right) \right. \nonumber \\
&+ \left.  e^{2 \left(r - x\right)/a_{0}} \Phi\left(-\frac{r - x}{\sigma} - \frac{2 \sigma}{a_{0}}\right) \right],
\label{eq:ElectronDensity_Eq_3}
\end{eqnarray}
where $\Phi\left(z\right) = \frac{1}{\sqrt{2 \pi}} \int^{z}_{-\infty} dt e^{- t^{2} / 2}$ is the cumulative distribution function of the Gaussian. We depict the densities of Eqs. \ref{eq:ElectronDensity_Eq_1} and \ref{eq:ElectronDensity_Eq_3} with $x = 0$ in Fig. \ref{fig2}(\textbf{a}) together with spherically averaged full-BO radial densities around hydrogen atoms in LiH. The covariance value used, $\langle \hat{u}^{2}_{\alpha H} \rangle = \sigma^{2} = 0.079$~au$^{2}$, is the computed covariance value of hydrogen in LiH at 293~K. Here we can see exactly the same phenomena in our simple model and Gaussian approach results for all electron density in LiH. We note that there is a slightly lower strict-BO and full-BO densities in our hydrogen like model system. This can be explained by the nature of the LiH crystal, where hydrogen receives some of the density from lithium, see the inset of Fig. \ref{fig2}(\textbf{b}).

An approximate form of Eq. \ref{eq:ElectronDensity_Eq_2} can be obtained by Taylor expanding $n_{R}\left(r\right) = n_{x + u}\left(r\right)$ in $u$ about $x$, namely
\begin{equation} 
n_{x + u}\left(r\right) \approx n_{x}\left(r\right) + \frac{dn_{x}\left(r\right)}{dx} u + \frac{1}{2} \frac{d^{2}n_{x}\left(r\right)}{dx^{2}} u^2,
\label{eq:ElectronDensity_Eq_4}
\end{equation}
where
\begin{eqnarray} 
\frac{dn_{x}\left(r\right)}{dx} &= \frac{2}{a_{0}} n_{x}\left(r\right) sgn\left(r - x\right), \nonumber \\
\frac{d^{2}n_{x}\left(r\right)}{dx^{2}} &= -\frac{4}{a_{0}} n_{x}\left(r\right) \delta\left(r - x\right).
\label{eq:ElectronDensity_Eq_5}
\end{eqnarray}
By using Eqs. \ref{eq:ElectronDensity_Eq_4} and \ref{eq:ElectronDensity_Eq_5} in Eq. \ref{eq:ElectronDensity_Eq_2}
\begin{equation} 
n\left(r\right) \approx n_{x}\left(r\right) - \frac{2 \sigma^{2} }{a_{0}} n_{x}\left(r\right) \delta\left(r - x\right).
\label{eq:ElectronDensity_Eq_6}
\end{equation}
We see that the second derivatives in the Taylor expansion lead to a discontinuity exactly at the equilibrium position and thus the Taylor polynomial approach cannot capture the correct electron density shape of Eq. \ref{eq:ElectronDensity_Eq_2} in this case. This is also the origin for the inability of the polynomial approach of Sec. \ref{HarmonicApproximationToDensity} to produce reliable all electron densities. However, the pseudo electron or valence electron densities do not in general have cusps at the nuclei equilibrium positions, but still care is needed in computing the finite difference derivatives of electron densities. We find that the bimodal shape persists as a function of displacement parameters used in the finite difference and the numerical derivatives appear accurate.

We have now the necessary pieces of information to draw a conclusion about the bimodal shapes discovered in the computed pseudo electron densities. The diagonal expansion $\left\langle n_{d}\right\rangle = n_{\vec{x}} + \frac{1}{2} \sum_{s} {\partial^{2}{n_{\vec{x}}}}/{\partial{x^{2}_{s}}} \left\langle \hat{u}^{2}_{s} \right\rangle$ is a rather accurate description of full-BO pseudo electron density, but  due to the reasons discussed here, the polynomial approach cannot be successfully used to describe the full-BO all electron density. In the diagonal expansion $\left\langle \hat{u}^{2}_{s} \right\rangle > 0$ and thus the reduction of electron density is caused by negative derivatives ${\partial^{2}{n_{\vec{x}}}}/{\partial{x^{2}_{s}}} < 0$ and thus $\frac{1}{2} \sum_{s} {\partial^{2}{n_{\vec{x}}}}/{\partial{x^{2}_{s}}} \left\langle \hat{u}^{2}_{s} \right\rangle < 0$. The finite order expansion does not accurately describe the densities at and near the nuclei equilibrium positions and increasing temperature makes the full-BO correction term more significant (still negative) causing the bimodal shape. Based on our findings, we therefore assess the bimodal shape of pseudo density in the polynomial approximation to be an artifact of the truncated expansion in the studied system.

We consider next the all electron densities and their temperature dependence shown in Fig. \ref{fig2}. The presented densities in LiH are averaged radial densities and in the case of hydrogen the model densities based on Eq. \ref{eq:ElectronDensity_Eq_3} are also shown. We find similar reduction in the case of all electron densities as in the case of pseudo electron densities. The relative changes are similar to the pseudo electron case, over 50\% in the case of hydrogen (Fig. \ref{fig2}(\textbf{a})) and over 80\% in the case of lithium (Fig. \ref{fig2}(\textbf{b})). We can see that the atomic density of hydrogen is lower than that of the averaged density of hydrogen in LiH, see Fig. \ref{fig2}(\textbf{a}). This can be explained by bonding in the crystalline phase where some of the electron density of lithium is transferred towards the hydrogen nuclei. We see similar reduction at the lithium equilibrium position in LiH, see the inset of Fig. \ref{fig2}(\textbf{b}). The nuclear one body densities are shown in Figs. \ref{fig2}(\textbf{a}) and \ref{fig2}(\textbf{b}) and width of the nuclear densities quite accurately match with the width of the radial electron density peaks. The temperature dependence of electron densities at the nuclear equilibrium positions is depicted in Fig. \ref{fig2}(\textbf{c}). The heavier nuclei have a stronger temperature dependence than hydrogen, which was also visible in Figs. \ref{fig1}(\textbf{d})-\ref{fig1}(\textbf{f}). This originates from the lower vibrational frequencies as these modes are occupied more rapidly as a function of temperature. This can also been seen through the behaviour of mean square displacements as a function of temperature, presented at the inset of Fig. \ref{fig2}(\textbf{c}). Namely, these quantities can be written as [Eq. \ref{eq:TheoryEq_7}] $\langle \hat{u}^{2}_{s} \rangle = { \hbar }/{ M_{s} } \sum_{j} \omega^{-1}_{j} e^{2}\left(s|j\right) \left(\bar{n}_{j} + {1}/{2}\right)$, where $\bar{n}_{j} = {1}/{(e^{\beta \hbar \omega_{j}} - 1)}$. The temperature dependence is only through $\bar{n}_{j}$ and thus for heavier masses causing the lower frequencies $\omega_{j}$ (see Fig. \ref{fig1}(\textbf{b})), we see a stronger temperature dependence in deuterium than in hydrogen.

\subsection{Comparison with experimental radial densities}
\label{ComparisonWithExperimentalRadialDensities}

\begin{figure}
\centering
\includegraphics[]{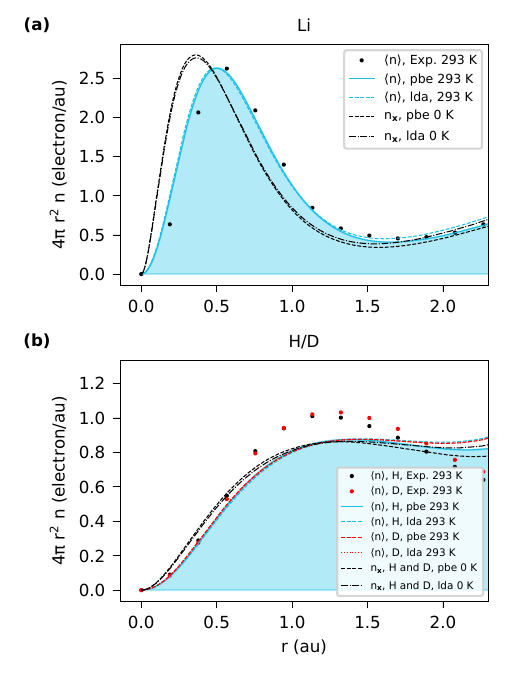}
\caption{Spherically averaged radial all electron densities in LiH. (\textbf{a}) For Li nucleus (\textbf{b}) for H and D nucleus. The experimental data at 293 K is from Ref. \cite{Vidal-EvidenceOnTheBreakdownOfTheBornOppenheimerApproximationInTheChargeDensityOfCrystalline7LiHD-1992}.}
\label{fig3}
\end{figure}
Our findings so far imply the breakdown of the strict BO approximation. We have verified the difference in densities of LiH and LiD in the volumes near H and D nuclei, which was suggested to be a sign of the BO breakdown in \cite{Vidal-EvidenceOnTheBreakdownOfTheBornOppenheimerApproximationInTheChargeDensityOfCrystalline7LiHD-1992}. Indeed, the electron densities of LiH and LiD are identical within the strict BO approximation, but are different already in the full BO approximation. Therefore the differences in the densities of LiH and LiD are not necessary a sign of the full BO breakdown. The measurements cannot be made within the BO approximation and thus are not, at least directly, able to distinct the possible BO breakdown near the Li nuclei. Our results show a significant reduction of electron density at the volumes near the nuclear equilibrium positions of all three species of elements.

For a comparison of experimental and theoretical electron densities, the computed spherically averaged radial densities with the previous experimental results at 293~K \cite{Vidal-EvidenceOnTheBreakdownOfTheBornOppenheimerApproximationInTheChargeDensityOfCrystalline7LiHD-1992} for LiH are depicted in Fig. \ref{fig3}. The radial strict BO density is flattened for both species by the full BO effects with quantum mechanical nuclei. For lithium, the computed values are rather close to the experimental ones at low and high radii ranges. The largest differences are at around $\sim 0.2-0.9$~au radius. The full BO densities, in comparison to the corresponding strict BO densities, are closer to the experimental values in most of the range considered. In the case of lithium, the radii range $\sim 1.0-2.2$~au, the experimental density compares relatively well to our computational result on full-BO densities. The results for hydrogen show that the smallest deviations are obtained at lowest radii and the largest deviation appears in the middle range. In this case the experimental results imply higher values at the middle range, which is somewhat the opposite to what we found for lithium. The largest difference between experimental and computational results at $r \approx 1.1$~au is around 20\%. Again, the strict BO densities are the furthest away from the experimental values, in particular at low radii range. We find the effect of the exchange-correlation functionals used (PBE/LDA) small in comparison to the effect due to the quantum mechanical nuclei in full-BO level. The experimental H–D differences in the radial densities are larger than those obtained in our calculations, and have been interpreted as evidence for a breakdown of the Born--Oppenheimer approximation \cite{Vidal-EvidenceOnTheBreakdownOfTheBornOppenheimerApproximationInTheChargeDensityOfCrystalline7LiHD-1992}. The largest differences are above $r \approx 1.2$~au radii. We find that the largest difference (around 20\%) between the theoretical and experimental values $r \approx 1.1$~au is much larger than the isotope effect.

\subsection{Comparison with experimental structure factors}
\label{ComparisonWithExperimentalStructureFactors}

Experimental and theoretical structure factors for LiH at 93K are listed in Table \ref{ComparisonWithExperimentalStructureFactorsTable_1}.
\begin{table*}
\caption{Experimental and theoretical structure factors for LiH at 93 K. The experimental data is from Table 3 of Ref. \cite{Vidal-EvidenceOnTheBreakdownOfTheBornOppenheimerApproximationInTheChargeDensityOfCrystalline7LiHD-1992}. We denote by $F^{exp}_{o}$ the observed structure-factor magnitudes and by $F^{exp}_{c}$ the calculated structure-factor magnitudes obtained from the refinement of the reference model of Ref. \cite{Vidal-EvidenceOnTheBreakdownOfTheBornOppenheimerApproximationInTheChargeDensityOfCrystalline7LiHD-1992}.}
\lineup
\centering
\begin{tabular}{@{}rrr rrrrrrrrr}
\br
$h$ & $k$ & $l$ & $F^{exp}_{o}$ & $F^{exp}_{c}$ & $\delta F^{exp}_{o}$ & $F^{pbe}_{c}$ & $F^{lda}_{c}$ & $F^{\vec{x}}_{c}$ & $\frac{F^{pbe}_{c}-F^{exp}_{o}}{F^{exp}_{o}}$ &$\frac{F^{\vec{x}}_{c}-F^{exp}_{o}}{F^{exp}_{o}}$ & $\frac{F^{\vec{x}}_{c}-F^{pbe}_{c}}{F^{pbe}_{c}}$ \\
\mr
0 & 0 & 0 & 16.0000 & 16.0000 & 0.0000 & 15.9989 & 15.9968 & 16.0043 & -0.0001 & 0.0003 & 0.0003 \\
1 & 1 & 0 & 4.2093 & 4.4277 & 0.0040 & 3.4328 & 3.3976 & 3.4907 & -0.1845 & -0.1707 & 0.0169 \\
2 & 0 & 0 & 8.4351 & 8.0910 & 0.0292 & 7.9802 & 7.7780 & 8.4177 & -0.0539 & -0.0021 & 0.0548 \\
2 & 2 & 0 & 6.0193 & 5.8829 & 0.0060 & 5.7736 & 5.5968 & 6.3571 & -0.0408 & 0.0561 & 0.1011 \\
3 & 1 & 1 & 4.0033 & 3.9911 & 0.0080 & 3.9374 & 3.8641 & 4.3099 & -0.0165 & 0.0766 & 0.0946 \\
2 & 2 & 2 & 4.5499 & 4.6269 & 0.0004 & 4.5464 & 4.3894 & 5.2104 & -0.0008 & 0.1452 & 0.1461 \\
4 & 0 & 0 & 3.8350 & 3.7645 & 0.0120 & 3.7058 & 3.5645 & 4.4381 & -0.0337 & 0.1573 & 0.1976 \\
3 & 3 & 1 & 2.8955 & 2.9264 & 0.0056 & 2.8829 & 2.7895 & 3.4239 & -0.0044 & 0.1825 & 0.1877 \\
4 & 2 & 0 & 3.1448 & 3.1237 & 0.0016 & 3.0771 & 2.9495 & 3.8270 & -0.0215 & 0.2169 & 0.2437 \\
4 & 2 & 2 & 2.5167 & 2.6281 & 0.0128 & 2.5909 & 2.4760 & 3.3516 & 0.0295 & 0.3317 & 0.2936 \\
3 & 3 & 3 & 2.1300 & 2.1634 & 0.0104 & 2.1325 & 2.0447 & 2.7442 & 0.0012 & 0.2883 & 0.2868 \\
5 & 1 & 1 & 2.1199 & 2.1634 & 0.0064 & 2.1341 & 2.0463 & 2.7441 & 0.0067 & 0.2944 & 0.2858 \\
4 & 4 & 0 & 1.8937 & 1.9177 & 0.0020 & 1.8928 & 1.7997 & 2.6542 & -0.0005 & 0.4016 & 0.4022 \\
5 & 3 & 1 & 1.6263 & 1.6302 & 0.0008 & 1.6108 & 1.5353 & 2.2449 & -0.0095 & 0.3804 & 0.3936 \\
4 & 4 & 2 & 1.6622 & 1.6576 & 0.0132 & 1.6368 & 1.5531 & 2.3893 & -0.0153 & 0.4374 & 0.4597 \\
6 & 0 & 0 & 1.6502 & 1.6576 & 0.0080 & 1.6373 & 1.5536 & 2.3904 & -0.0078 & 0.4485 & 0.4600 \\
6 & 2 & 0 & 1.4356 & 1.4417 & 0.0016 & 1.4246 & 1.3495 & 2.1655 & -0.0076 & 0.5084 & 0.5200 \\
5 & 3 & 3 & 1.2522 & 1.2505 & 0.0060 & 1.2386 & 1.1756 & 1.8720 & -0.0109 & 0.4950 & 0.5114 \\
6 & 2 & 2 & 1.2486 & 1.2607 & 0.0032 & 1.2466 & 1.1792 & 1.9727 & -0.0016 & 0.5799 & 0.5825 \\
4 & 4 & 4 & 1.0973 & 1.1076 & 0.0056 & 1.0952 & 1.0346 & 1.8057 & -0.0019 & 0.6456 & 0.6487 \\
7 & 1 & 1 & 0.9884 & 0.9741 & 0.0040 & 0.9679 & 0.9161 & 1.5863 & -0.0208 & 0.6049 & 0.6389 \\
5 & 5 & 1 & 0.9844 & 0.9741 & 0.0056 & 0.9677 & 0.9158 & 1.5864 & -0.0170 & 0.6116 & 0.6394 \\
6 & 4 & 0 & 0.9766 & 0.9771 & 0.0024 & 0.9680 & 0.9137 & 1.6604 & -0.0088 & 0.7002 & 0.7152 \\
6 & 4 & 2 & 0.8702 & 0.8652 & 0.0052 & 0.8583 & 0.8095 & 1.5327 & -0.0136 & 0.7613 & 0.7857 \\
5 & 5 & 3 & 0.7818 & 0.7689 & 0.0040 & 0.7667 & 0.7243 & 1.3626 & -0.0193 & 0.7429 & 0.7772 \\
7 & 3 & 1 & 0.7720 & 0.7689 & 0.0052 & 0.7669 & 0.7245 & 1.3627 & -0.0065 & 0.7651 & 0.7767 \\
8 & 0 & 0 & 0.7021 & 0.6849 & 0.0036 & 0.6820 & 0.6425 & 1.3194 & -0.0287 & 0.8792 & 0.9347 \\
\br
\end{tabular}
\label{ComparisonWithExperimentalStructureFactorsTable_1}
\end{table*}
We give the full-BO structure factors computed with PBE ($F^{pbe}_{c}$) and LDA ($F^{lda}_{c}$). In addition the strict BO structure factors computed with PBE ($F^{\vec{x}}_{c}$) are listed. In our notation $F\left(\vec{G}\right) = \int_{cell} d\vec{y} n\left(\vec{y}\right) e^{- i \vec{G} \cdot \vec{y}}$ and $F^{pbe/lda}_{c} = \left| F\left(\vec{G}\right) \right|$. In the case of full-BO densities, the largest difference between the experiments is at $\left(h,k,l\right) = (1, 1, 0)$ and the relative difference $\frac{F^{pbe}_{c}-F^{exp}_{o}}{F^{exp}_{o}}$ is around -18\%. At this particular point, the strict BO density is more accurate with $\frac{F^{\vec{x}}_{c}-F^{exp}_{o}}{F^{exp}_{o}} \approx -17$\%. At the remaining values $\left(h,k,l\right)$, the full-BO structure factor relative difference $\frac{F^{pbe}_{c}-F^{exp}_{o}}{F^{exp}_{o}} \approx -1.2$\%, on average. For the strict BO structure factors the corresponding number is $\frac{F^{\vec{x}}_{c}-F^{exp}_{o}}{F^{exp}_{o}} \approx 42$\%. The average over all the relative differences $\frac{F^{pbe}_{c}-F^{lda}_{c}}{F^{lda}_{c}}$ from the values given in Table \ref{ComparisonWithExperimentalStructureFactorsTable_1} is around 4.5 \%.  A particular trend in strict-BO structure factors is the increasing discrepancy between the experiment for larger Miller indices. In contrast, the full-BO structure factors have similar differences to experimental values at all $\left(h,k,l\right)$ outside $(1, 1, 0)$. The significant difference originates from the inability of strict-BO approximation to capture the accurate electron density at the vicinity of the nuclei equilibrium positions. Interestingly, the 18\% discrepancy of the $(1,1,0)$ structure-factor magnitude does not contribute significantly to the observed 20\% difference between the experimental and theoretical radial densities at $r \approx 1.1$~au visible in Fig.~\ref{fig3}. This follows from the Fourier relation between $n\left(\vec{y}\right)$ and the structure factors: at this radius the $(1,1,0)$ reflection accounts for only about 15\% of the relevant Fourier weight, so even an 18\% change in $|F^{pbe}_{c,110}|$ modifies the spherically averaged radial density by only a few percent. The much larger $\sim 20\%$ deviation in Fig.~\ref{fig3} cannot be likely explained by this structure factor discrepancy alone. We provide additional experimental and theoretical structure factors in Table~\ref{StructureFactorsTable_1} of \ref{StructureFactors}, where the discrepancy at $(1,1,0)$ is seen to persist at all temperatures, also in the case of LiD.

\section{Discussion and conclusions}
\label{DiscussionAndConclusions}

We have studied corrections beyond the strict BO approximation to the electronic structure of LiH and LiD and find a significant failure of the strict BO approximation, particularly in regions near the nuclei. The temperature dependence of electron density was found to be stronger at the vicinity of the heavier nuclei. While our computational results compare rather well to the earlier experimental results, there still remains a room for improvement. There are several factors that can cause the mentioned differences in electron densities near hydrogen atoms and in the computational and experimental structure factors in particular at $\left(1,1,0\right)$. Possible factors are errors originating from the effects beyond the full BO approximation, from the reconstruction of all electron density from PAW datasets and the neglected of anharmonicity. In addition, the experimental X-ray structure factors for hydrogen are known to be among the most difficult to determine reliably of all the elements  \cite{Noritake-ChargeDensityMeasurementInMgH2bySynchrotronXrayDiffraction-2003,Jha-TAAMaReliableAndUserFriendlyToolForHydrogenAtomLocationUsingRoutineXrayDiffractionData-2020}. We note, however, that the computed phonon dispersions here are consistent with earlier computational results, which appear to closely match the experimental findings \cite{Zhang-PhononAndElasticInstabilitiesInRocksaltAlkaliHydridesUnderPressureFirstPrinciplesStudy-PhysRevB.75.104115-2007}. Moreover, our computed mean square displacement value, $\langle \hat{\vec{u}}^{2}_{s} \rangle$, for hydrogen is $0.217$~au$^{2}$ at 20 K while the corresponding experimental value \cite{Colognesi-ExtractionOfTheDensityOfPhononStatesInLiHandNaH-2004} is $0.199-0.221$~au$^{2}$, depending on the method of measuring. The computed value was obtained with $8 \times 8 \times 8$ $\vec{q}$ point grid while the smaller $2 \times 2 \times 2$ $\vec{q}$ point grid gave the value $0.215$~au$^{2}$ at 20 K. Both these findings, the phonon dispersion matching to experiments and the value of the mean square displacement, imply a relatively weak anharmonicity. The $\sim 18\%$ discrepancy between the experimental and theoretical structure factors at $(1,1,0)$ in LiH and LiD represents an interesting open issue for future work.

Earlier computational results on YH$_{6}$ superconductor and Cs-IV phase of hydrogen with the polynomial approach \cite{Harkonen-BreakdownOfTheBornOppenheimerApproximationInSolidHydrogenAndHydrogenRichSolids-Arxiv-2023} imply the failure of the strict BO approximation in phases of matter that exist at high pressures. Here our results show that the strict BO approximation can also fail in states of matter that exist at low pressures, and that the dominant mechanism of the breakdown appears to be associated with the local position uncertainty of the nuclei. Another important aspect indicated by the results is that a significant beyond strict BO effects can occur in elements other than hydrogen, lithium and deuterium in the present case. The lithium is around seven times more massive than hydrogen which suggests that there could be relevant full BO effects in a number of different solids. For instance, carbon has less than twice the mass of lithium.

To summarize. We report a significant breakdown of the strict BO approximation in LiH and LiD, which is verified by computing the ab-initio full BO electron density with density functional methods. Our results partially explain the earlier experimental findings \cite{Vidal-EvidenceOnTheBreakdownOfTheBornOppenheimerApproximationInTheChargeDensityOfCrystalline7LiHD-1992} and support the earlier computational findings in solid hydrogen and in YH$_{6}$ superconductor \cite{Harkonen-BreakdownOfTheBornOppenheimerApproximationInSolidHydrogenAndHydrogenRichSolids-Arxiv-2023}. Our recent findings highlight the importance of effects beyond the strict BO in solids, which are likely necessary to consider in order to enhance our understanding of materials such as various hydrides and solid hydrogen.

\section{Data availability statement}
\label{DataAvailabilityStatement}

The input and output data used in this work, are openly available in Zenodo under the title ''On the breakdown of the Born-Oppenheimer approximation in LiH and LiD:
data and scripts'' at DOI: https://doi.org/10.5281/zenodo.17953036.

\section{Acknowledgments}
\label{Acknowledgments}

We gratefully acknowledge funding from the Magnus Ehrnrooth foundation and Jenny and Antti Wihuri foundation. We acknowledge Prof. E. K. U. Gross for numerous discussions on physics related to this work over the years. We also thank Msc. Aleksi Hartikainen for discussions related to the results of this work. The computing resources for this work were provided by CSC - the Finnish IT Center for Science.

\appendix
\section{Structure Factors}
\label{StructureFactors}

The theoretical and experimental structure factors at 160 K and 293K are listed in Table \ref{StructureFactorsTable_1}. The used notation is described in the main text.
\begin{table*}
\caption{Experimental and theoretical structure factors for LiH and LiD at 160 K and 293K. The experimental data is from Table 3 of Ref. \cite{Vidal-EvidenceOnTheBreakdownOfTheBornOppenheimerApproximationInTheChargeDensityOfCrystalline7LiHD-1992}.}
\lineup
\centering
\begin{tabular}{@{}rrr rrrrrrrr}
\br
$h$ & $k$ & $l$ & $F^{exp,LiH}_{o, 160K}$ & $F^{pbe,LiH}_{c, 160K}$ & $F^{exp, LiH}_{o, 293K}$ & $F^{pbe,LiH}_{c, 293K}$ &  $F^{exp,LiD}_{o, 160K}$ & $F^{pbe,LiD}_{c, 160K}$ &  $F^{exp,LiD}_{o, 293K}$ & $F^{pbe,LiD}_{c, 293K}$ \\
\mr
0 & 0 & 0 & 16.0000 & 15.9988 & 16.0000 & 15.9986 & 16.0000 & 15.9996 & 16.0000 & 15.9986 \\
1 & 1 & 0 & 4.1756 & 3.4443 & 4.0588 & 3.4811 & 4.3689 & 3.4728 & 4.2882 & 3.4811 \\
2 & 0 & 0 & 8.3870 & 7.9410 & 8.2616 & 7.8126 & 8.1853 & 7.9889 & 8.0413 & 7.8126 \\
2 & 2 & 0 & 5.9380 & 5.7126 & 5.7298 & 5.5175 & 5.9344 & 5.7528 & 5.7133 & 5.5175 \\
3 & 1 & 1 & 3.9126 & 3.8697 & 3.6458 & 3.6610 & 3.9572 & 3.8354 & 3.7229 & 3.6610 \\
2 & 2 & 2 & 4.4478 & 4.4723 & 4.1695 & 4.2394 & 4.6041 & 4.5054 & 4.3396 & 4.2394 \\
4 & 0 & 0 & 3.7482 & 3.6243 & 3.4444 & 3.3718 & 3.7854 & 3.6517 & 3.4851 & 3.3718 \\
3 & 3 & 1 & 2.7913 & 2.8009 & 2.4665 & 2.5526 & 2.8830 & 2.7776 & 2.6080 & 2.5526 \\
4 & 2 & 0 & 3.0080 & 2.9919 & 2.6804 & 2.7315 & 3.0848 & 3.0146 & 2.7886 & 2.7315 \\
4 & 2 & 2 & 2.5097 & 2.5045 & 2.1712 & 2.2438 & 2.5860 & 2.5234 & 2.2781 & 2.2438 \\
3 & 3 & 3 & 2.0275 & 2.0480 & 1.7330 & 1.7978 & 2.1042 & 2.0319 & 1.8160 & 1.7978 \\
5 & 1 & 1 & 2.0255 & 2.0495 & 1.7275 & 1.7993 & 2.0984 & 2.0334 & 1.8101 & 1.7993 \\
4 & 4 & 0 & 1.7885 & 1.8083 & 1.4982 & 1.5600 & 1.8876 & 1.8218 & 1.5950 & 1.5600 \\
5 & 3 & 1 & 1.5247 & 1.5292 & 1.2489 & 1.2932 & 1.6048 & 1.5178 & 1.3357 & 1.2932 \\
4 & 4 & 2 & 1.5377 & 1.5546 & 1.2592 & 1.3160 & 1.6240 & 1.5660 & 1.3523 & 1.3160 \\
6 & 0 & 0 & 1.5407 & 1.5551 & 1.2503 & 1.3165 & 1.6227 & 1.5665 & 1.3582 & 1.3165 \\
6 & 2 & 0 & 1.3265 & 1.3452 & 1.0655 & 1.1175 & 1.4242 & 1.3550 & 1.1473 & 1.1175 \\
5 & 3 & 3 & 1.1505 & 1.1623 & 0.8970 & 0.9467 & 1.2238 & 1.1540 & 0.9722 & 0.9467 \\
6 & 2 & 2 & 1.1656 & 1.1702 & 0.9094 & 0.9540 & 1.2444 & 1.1787 & 0.9927 & 0.9540 \\
7 & 0 & 1 & - & 0.8845 & 0.7481 & 0.8756 & - & 0.8911 & - & 0.8756 \\
7 & 1 & 1 & 0.8881 & 0.8978 & 0.6646 & 0.7044 & 0.9542 & 0.8917 & 0.7360 & 0.7044 \\
5 & 5 & 1 & 0.8932 & 0.8976 & 0.6607 & 0.7042 & 0.9526 & 0.8915 & 0.7297 & 0.7042 \\
6 & 4 & 0 & 0.8889 & 0.8982 & 0.6607 & 0.7051 & 0.9556 & 0.9045 & 0.7315 & 0.7051 \\
6 & 4 & 2 & 0.7825 & 0.7917 & 0.5688 & 0.6099 & 0.8513 & 0.7972 & 0.6300 & 0.6099 \\
5 & 5 & 3 & 0.6984 & 0.7030 & 0.4843 & 0.5312 & 0.7481 & 0.6984 & 0.5439 & 0.5312 \\
7 & 3 & 1 & 0.7006 & 0.7032 & 0.4949 & 0.5314 & 0.7490 & 0.6986 & 0.5500 & 0.5314 \\
8 & 0 & 0 & 0.6152 & 0.6217 & 0.4280 & 0.4612 & 0.6706 & 0.6259 & 0.4837 & 0.4612 \\
\br
\end{tabular}
\label{StructureFactorsTable_1}
\end{table*}

\section*{ORCID iDs}

Ville H\"{a}rk\"{o}nen: https://orcid.org/0000-0002-2956-5457

\section*{References}
\bibliography{bibfile}
\end{document}